\documentclass[pdflatex,sn-nature]{sn-jnl}

\usepackage{graphicx}
\usepackage{multirow}
\usepackage{amsmath,amssymb,amsfonts}
\usepackage{amsthm}
\usepackage{mathrsfs}
\usepackage[title]{appendix}
\usepackage{xcolor}
\usepackage{textcomp}
\usepackage{manyfoot}
\usepackage{booktabs}
\usepackage{algorithm}
\usepackage{algorithmicx}
\usepackage{algpseudocode}
\usepackage{listings}
\usepackage[normalem]{ulem}

\hypersetup{hypertexnames=false}

\begin{document}

\title[TSBench]{TSBench: A physics-grounded benchmark for evaluating LLM understanding of chemical reaction mechanisms}

\author[1,2]{\fnm{Xiaohu} \sur{Xu}}

\author*[1,2,3]{\fnm{Tong} \sur{Zhu}}\email{tzhu@lps.ecnu.edu.cn}

\affil[1]{\orgdiv{Shanghai Engineering Research Center of Molecular Therapeutics and New Drug Development, School of Chemistry and Molecular Engineering}, \orgname{East China Normal University}, \orgaddress{\city{Shanghai}, \postcode{200062}, \country{P. R. China}}}

\affil[2]{\orgname{Shanghai Innovation Institute}, \orgaddress{\city{Shanghai}, \postcode{200003}, \country{P. R. China}}}

\affil[3]{\orgname{NYU-ECNU Center for Computational Chemistry at NYU Shanghai}, \orgaddress{\city{Shanghai}, \postcode{200062}, \country{P. R. China}}}

\abstract{Understanding a chemical reaction requires mapping a symbolic reactant--product description onto the three-dimensional pathway through which atoms rearrange, yet chemistry benchmarks for large language models (LLMs) largely probe factual knowledge and text-based reasoning. Here we introduce TSBench, a benchmark in which an LLM agent uses structure-editing tools to construct three-dimensional transition-state (TS) guesses verified by an automated quantum-chemical pipeline, yielding a physics-grounded pass/fail verdict. Across 546 evaluations of seven frontier LLMs on 78 elementary reactions, the aggregate success rate rose from 50.4\% to 66.8\% under diagnosis-driven revision; the best models approached 90\% on the simplest reactions, yet performance dropped sharply with mechanistic complexity. Most failed attempts produced locally plausible saddle points whose reaction paths led to the wrong reactant--product pair, revealing local geometric intuition without a robust grasp of the global reaction coordinate. TSBench establishes a mechanism-level yardstick for LLM agents in mechanism-sensitive tasks such as synthesis planning and autonomous experimentation.}


\maketitle

\section{Introduction}\label{sec:introduction}

Understanding a chemical reaction requires more than knowing which bonds break
and form: it requires reasoning about how atoms rearrange in three-dimensional
space to traverse the energy barrier separating reactants from products. This
mapping from a symbolic reaction description to a physical, three-dimensional
process is what distinguishes mechanistic understanding from factual recall and
underpins a wide range of predictive tasks in chemistry---from reaction outcome
prediction and retrosynthetic planning to reaction-condition optimisation and
self-driving laboratories~\cite{Coley2017_ReactionPrediction,Segler2018_SynthesisPlanning,Schwaller2019_MolecularTransformer,Shields2021_BayesianReactionOptimization,White2023_FutureChemistryLanguage,Tom2024_SelfDrivingLabs}. It is also the
kind of reasoning that is hardest to evaluate, because a mechanistic hypothesis ultimately has to be tested against physical criteria rather than only compared with a text answer key.

Large language models (LLMs) are increasingly being asked to perform exactly
this kind of reasoning. They have moved rapidly from general-purpose text
systems to chemistry-specific assistants that predict molecular and reaction
properties~\cite{Jablonka2024_GPT3,Zhang2025_Chemma,Zheng2025_MolecularPropertyDiscovery,Wellawatte2025_XpertAI}, plan
syntheses and elucidate mechanisms by guiding traditional search
algorithms~\cite{Bran2026_ChemicalReasoning}, and orchestrate computational and
experimental tools as autonomous chemistry
agents~\cite{Bran2024_ChemCrow,Boiko2023_Coscientist,Ruan2024_LLMRDF,Ramos2025_LLMReviewChemistry}. Related systems now span scientific-discovery loops,
AI-assisted synthesis, literature-to-robot execution, reaction-data mining and
computational workflow automation~\cite{Gottweis2026_CoScientist,Ghareeb2026_Robin,Li2026_MOSAIC,Pagel2026_ChemputationLLM,Li2026_ReactionSeek,Pham2026_ChemGraph,Soleymanibrojeni2026_GENIUS}. As these models are increasingly proposed as components of self-driving
laboratories and agentic synthesis platforms, the question is shifting from whether
they can retrieve and articulate chemical knowledge to whether they can apply
that knowledge to solve well-defined chemical problems, such as inferring
reaction pathways, planning syntheses and coordinating computational or
experimental workflows. Reaction mechanism understanding provides a
particularly consequential test of this capability: synthesis planning,
catalyst screening and process design all depend on correctly reasoning about
how a reaction proceeds.

A growing number of benchmarks has begun to evaluate LLMs in chemistry, but along
axes that leave reaction mechanism understanding largely unmeasured.
Early chemistry evaluations and question-answering benchmarks scored against expert
chemists~\cite{CastroNascimento2023_ChatGPTChemistry,White2023_CodeChemistry,Guo2023_ChemLLMBench,Mirza2025_ChemBench,Runcie2026_ChemIQ} and quantitative chemistry
tests~\cite{Xie2025_QCBench} probe factual knowledge and numerical problem
solving; multi-hop~\cite{Khodadad2026_ChemComp} and
modular-operation benchmarks~\cite{Li2025_ChemCoTBench,Ko2025_ReactionReasoner}
test stepwise reasoning over molecular transformations; multimodal and spatial
benchmarks~\cite{Alampara2025_MaCBench,Lv2026_AtomWorld,Cui2025_ChemOlympiad,Laghuvarapu2025_MolTextQA} test interpretation
and manipulation of chemical images, molecular structures and atomic structures.
Related tests further probe representation robustness, error correction,
fine-tuning for chemistry and materials tasks, and tool use in materials
science~\cite{Ganeeva2025_SMILESRobustness,Wu2025_MolErr2Fix,VanHerck2025_FineTunedChemMaterials,Liu2025_MatTools}. Together these
evaluations show that current models are competitive on chemical knowledge yet
uneven on reasoning, spatial integration and multi-step inference. Critically,
however, few of them require the model to translate a symbolic reaction
description into a concrete three-dimensional configuration whose physical
validity can be independently verified---the very capability that distinguishes
genuine mechanistic understanding from pattern matching over reaction databases.

The transition state (TS) offers a natural, physics-grounded probe of this
capability. A valid TS is a first-order saddle point on the potential energy
surface whose intrinsic reaction coordinate (IRC) descends to the specified
reactant and product endpoints~\cite{Fukui1981_IRC}---criteria that can be checked by quantum-chemical
calculation rather than by text matching. Constructing a TS from reactant and
product information forces a model to reason about how atoms are arranged at the
saddle point, how surrounding groups rotate or translate to bring reactive
centres into contact, and whether the resulting structure connects the intended
endpoints along a single reaction coordinate. Recent automated TS-search and
reaction-path exploration methods
have substantially reduced the manual effort required to locate transition
states~\cite{Zhang2025_RDA,Lee2025_ReactionGraphTS,Goodfellow2026_GraphRC,Kikkawa2025_DRLTS,Chen2025_LLMPathwayExploration,Zev2026_RxnNet,Madhavan2026_TSAgent}, but
these methods characterise transition states once a reaction is encoded in their
framework; they do not evaluate whether an LLM agent can itself formulate the
mechanistic hypothesis that turns reactant--product information into a
physically viable TS guess. Rather than competing with these specialised search algorithms, TSBench aims to
use the quantum-chemical validation pipeline that underlies them as a
physical evaluator of an LLM agent's mechanistic reasoning.

Here we introduce TSBench, a benchmark that evaluates LLM agents on reaction
mechanism understanding by requiring them to construct three-dimensional TS
structures from reactant and product information using a small set of
structure-editing and geometry-querying tools. Each proposed structure is
verified by an automated quantum-chemical pipeline---TS optimisation, frequency
analysis, IRC propagation and endpoint matching---that returns an objective
pass/fail verdict together with the diagnostic step at which the candidate
fails. Failed candidates are returned to the agent for revision, creating a
closed-loop generate--validate--diagnose--revise cycle that resembles the iterative workflow used by computational chemists when refining TS guesses.
The complete evaluation workflow is summarised in Fig.~\ref{fig:tsbench_workflow}.

We evaluate seven frontier LLMs on 78 elementary reactions distributed across
three subsets---Simple, Intermediate and Complex elementary reactions, containing 19, 40 and 19 reactions, respectively--- giving 546 model--reaction
evaluations. The aggregate success rate rose from 50.4\% after the initial generation round to 66.8\% after two rounds of diagnosis-driven revision, with the best models approaching
90\% success on the simplest reactions but dropping sharply as mechanistic
complexity increased. The dominant failure mode was endpoint mismatch: models
routinely produced plausible saddle points whose IRC connected the wrong
chemical endpoints---indicating that current models can acquire local geometric
intuition about saddle-point topology yet lack a robust grasp of the global
reaction coordinate. By coupling agentic structure construction to
quantum-chemical validation, TSBench moves chemistry LLM evaluation beyond
knowledge retrieval toward a physics-grounded, mechanism-level measure of
reaction understanding, and it exposes a specific, quantifiable gap---the
disconnect between local geometric plausibility and global mechanistic
correctness---that must be closed before LLM agents can be trusted with mechanism-sensitive reaction tasks.

\section{Results and discussion}\label{sec:results}

\subsection{TSBench task and evaluation setup}

TSBench probes mechanism-level reasoning by asking an LLM agent to convert a reactant--product specification into a three-dimensional TS candidate that passes quantum-chemical validation (Fig.~\ref{fig:tsbench_workflow}). To solve the task, the agent must infer how the reacting atoms approach one another, how the surrounding groups should be arranged, and how far the structure should progress along the reaction coordinate. The validator then provides an objective, physics-grounded verdict: a candidate passes only if the optimised saddle point connects the target reactant--product pair through bidirectional IRC calculations.

We applied this protocol to 78 elementary reactions retained from three published reaction resources after xTB-based compatibility screening: a compact small-molecule reaction set~\cite{Zhang2025_RDA}, the textbook elementary-reaction collection~\cite{Li2025_MDCDNN} and the BH9 benchmark~\cite{Prasad2022_BH9}. These retained reactions define the Simple, Intermediate and Complex elementary regimes summarised in Fig.~\ref{fig:dataset_overview}; detailed corpus construction and xTB screening procedures are described in Methods.

We conducted one benchmark evaluation for every combination of seven frontier models and 78 reactions, giving 546 model--reaction evaluations. We analyse the resulting records at three levels: subset-resolved performance across the three regimes, round-wise gains from diagnosis-driven revision, and failure-stage decomposition of unsuccessful searches.

\subsection{Performance landscape across chemical regimes}\label{sec:results_landscape}

\begin{figure}[t]
    \centering
    \begin{minipage}{90mm}
        \centering
        \includegraphics[width=\linewidth,keepaspectratio]{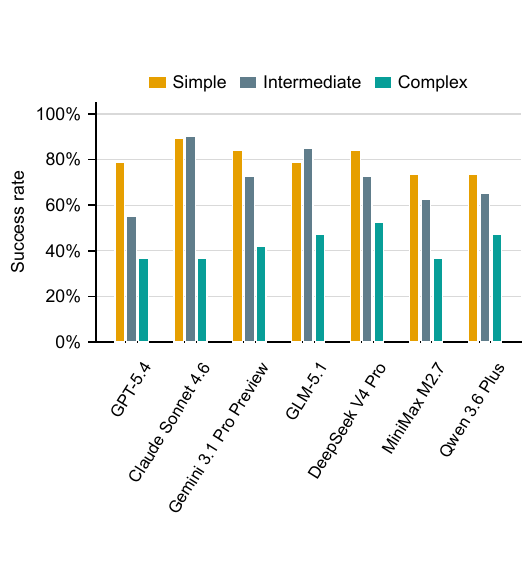}
        \caption{\textbf{Three-round TS-search success across reaction regimes.}
        Bars report the fraction of model--reaction evaluations that achieved strict
        success by the end of the three-round budget, broken down by model and
        the Simple elementary reactions, Intermediate elementary reactions and
        Complex elementary reactions subsets. Success on the Complex elementary subset
        remains below 53\% for all seven models.}
        \label{fig:success_rate}
    \end{minipage}
\end{figure}

Across the seven evaluated models, TSBench performance is organised by a clear
chemical-regime gradient (Fig.~\ref{fig:success_rate}). On the Simple elementary subset, all seven models locate a validated TS for more than 70\% of reactions within the three-round budget: Claude Sonnet 4.6
reaches 89.5\%, Gemini 3.1 Pro Preview and DeepSeek V4 Pro both reach
84.2\%, and the inter-model spread is 15.8 percentage points. On the
Intermediate elementary subset, the leading models retain comparable headline
success (Claude Sonnet 4.6 at 90.0\%; GLM-5.1 at 85.0\%), but the spread
widens to 35.0 points, with GPT-5.4 trailing at 55.0\%. On the Complex
elementary subset, the spread narrows again to 15.8 points, but now at a much
lower ceiling: success rates stay below 53\% for all seven models, with DeepSeek
V4 Pro setting the upper bound at 52.6\%. The three regimes therefore play complementary diagnostic roles: Simple elementary reactions are broadly
accessible to current agents, Intermediate elementary reactions separate model
capabilities most strongly, and Complex elementary reactions expose a shared
performance ceiling.

Figure~\ref{fig:dataset_overview}a,b provides a quantitative view of the
structural differences between the three regimes. The Simple elementary subset
occupies the smallest heavy-atom range and contains few ring-containing cases;
the Intermediate elementary subset extends to larger organic systems; and the
Complex elementary subset contains the broadest heavy-atom distribution and the
highest ring-count cases. Although heavy-atom and ring counts do not by
themselves define mechanistic difficulty, they show that the three regimes
differ systematically in molecular size and topological constraint. The
performance gradient in Fig.~\ref{fig:success_rate} therefore occurs alongside
a measurable increase in the structural context surrounding each TS-search
task.

\subsection{Diagnosis-driven revision gains across rounds}\label{sec:results_revision}

\begin{figure}[t]
    \centering
    \begin{minipage}{90mm}
        \centering
        \includegraphics[width=\linewidth,keepaspectratio]{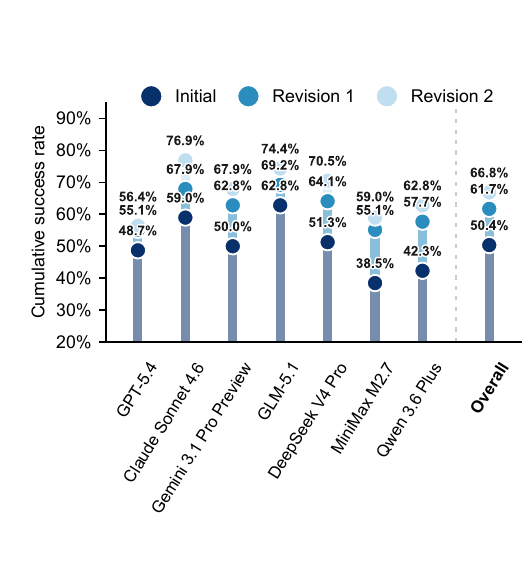}%
        \caption{\textbf{Cumulative TS-search success across revision rounds.}
        For each model, the three markers show the cumulative strict-success rate
        after the initial generation (Initial) and after the first and second
        diagnosis-driven revision rounds (Revision 1 and Revision 2).
        Success Rates are calculated over the same 78 reactions for each model, 
        comprising 19 Simple, 40 Intermediate and 19 Complex elementary reactions; 
        the rightmost \emph{Overall} column pools all 546 model--reaction evaluations.}
        \label{fig:cumulative_overall}
    \end{minipage}
\end{figure}

Diagnosis-driven revision substantially improves TS-search success across the
benchmark (Fig.~\ref{fig:cumulative_overall}). Pooled across the 546
model--reaction evaluations, the cumulative strict-success rate rises from
50.4\% after the initial generation round to 61.7\% after one revision
round and 66.8\% after two revision rounds, corresponding to 275, 337 and
365 successful evaluations. The 16.4 points absolute gain reduces
the unresolved fraction from 49.6\% to 33.2\%, showing that the diagnostic
loop recovers a substantial fraction of initially failing searches.

The gains are concentrated in the first diagnostic revision. Six of seven
models record their largest jump in the first revision round (between 6.4 and 16.7 points) and
smaller gains in the second (between 1.3 and 6.4 points); Claude Sonnet 4.6 is the
lone exception, recovering seven additional reactions in each revision round (9.0 points per round, 17.9 points in total); relative to its 32 initially unresolved evaluations, this is the largest recovered share in the cohort (43.8\%, versus 15.0--39.5\% for the other models). At the population
level, the 16.4-point gain splits into 11.3 points from the first revision and
5.1 points from the second. The subset-resolved cumulative tables
(Supplementary Tables~S1--S3) show that both revision rounds contribute gains
on each of the three subsets, but the Complex elementary subset remains below
53\% cumulative success even after two revisions. Thus, revision improves
outcomes across the benchmark without removing the low-ceiling behaviour on
the most demanding subset.

Successful revision often involves replacing the proposed mechanism rather
than only refining local geometry. Within the three-round budget, 2703 of 3378
classifiable diagnostic verdicts (80.0\%) call for a new mechanism (Supplementary
Fig.~S7). Of the 90 evaluations that first succeeded in a revision round, 85
followed a classifiable diagnostic verdict, and 71 of these (83.5\%) succeeded
after a ``replace the mechanism'' verdict in the previous round (Supplementary
Fig.~S8). These
statistics indicate that the diagnostic loop frequently contributes by
redirecting the generator from a failed mechanistic hypothesis to a new TS
candidate, rather than by simply polishing the previous structure.

\subsection{Failed searches often follow the wrong reaction coordinate}\label{sec:results_failure}

\begin{figure}[t]
    \centering
    \begin{minipage}{90mm}
        \centering
        \includegraphics[width=\linewidth,keepaspectratio]{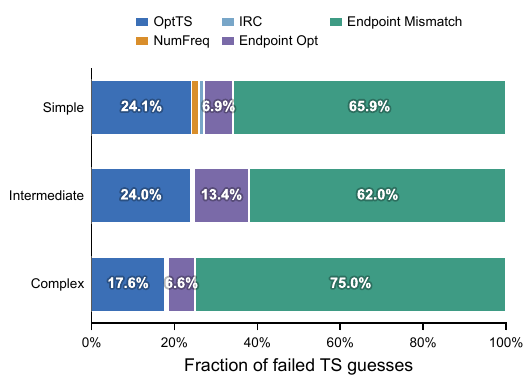}%
        \caption{\textbf{Failure-mode composition of TS guesses across reaction subsets.}
        Each horizontal bar shows the distribution of failure stages, normalised
        to 100\% within each of the Simple elementary reactions, Intermediate
        elementary reactions and Complex elementary reactions subsets, for guesses
        that did not achieve strict success. Stages are ordered along the validation pipeline: TS optimisation (OptTS), frequency analysis (NumFreq), IRC propagation (IRC), endpoint optimisation (Endpoint Opt) and endpoint matching (Endpoint Mismatch). Endpoint Mismatch
        (IRC endpoints not matching the target reactant--product pair) is the dominant category in all three
        subsets; TS-optimisation failures appear as a secondary contribution
        across all three subsets.}
        \label{fig:failure_steps}
    \end{minipage}
\end{figure}

Across the three subsets, failed TS guesses are dominated by a single
signature: optimised TS candidates that pass the preceding validation stages
but whose IRC endpoints do not match the target reactant--product pair
(Fig.~\ref{fig:failure_steps}, Supplementary Table~S4). Endpoint mismatch
accounts for 1137 of 1516 failed guesses (75.0\%) on the Complex elementary
subset, 1374 of 2217 (62.0\%) on Intermediate elementary and 579 of 879
(65.9\%) on Simple elementary. 
% \blueadd{Guesses that failed during input preparation (25 across the benchmark) are excluded from these counts (Supplementary Table~S4).} 
TS-optimisation failures, the next largest
category, contribute less on the Complex elementary subset (17.6\%) than on
Simple elementary or Intermediate elementary (24.1\% and 24.0\%).
Frequency, IRC and endpoint-optimisation errors together account for the
remaining 7.4--14.1\% of failed guesses. The pattern extends to the strongest agents: on the Complex elementary subset, 82.3\% of Claude Sonnet
4.6's failures are endpoint mismatch, compared with 66.4\% for MiniMax-M2.7
(Supplementary Table~S5).

At the model--reaction level the same pattern holds for terminal failures.
All 181 unresolved evaluations had a determinable terminal failure stage:
131 (72.4\%) terminated with endpoint mismatch, 40 (22.1\%) with
TS-optimisation failure and 10 (5.5\%) with endpoint-optimisation issues. Endpoint
mismatch therefore persists as the dominant failure mode through the
three-round budget: even after two diagnostic revisions, roughly three of four
residual failures still terminate with endpoints that do not match the target
reaction.

This pattern marks the central capability boundary identified by TSBench. In
many failed searches, the model can produce a local structure with the expected
features of a transition state, but the structure still lies on a reaction
coordinate that does not connect the target reactant--product pair. Current
models therefore show a partial grasp of what a TS should look like locally,
while still lacking robust understanding of the full reaction mechanism or
global reaction coordinate for each specific reaction.

\subsection{Case study: mechanism replacement under diagnostic feedback}\label{sec:results_case_study}

\begin{figure*}[t]
    \centering
    \includegraphics[width=\textwidth,keepaspectratio]{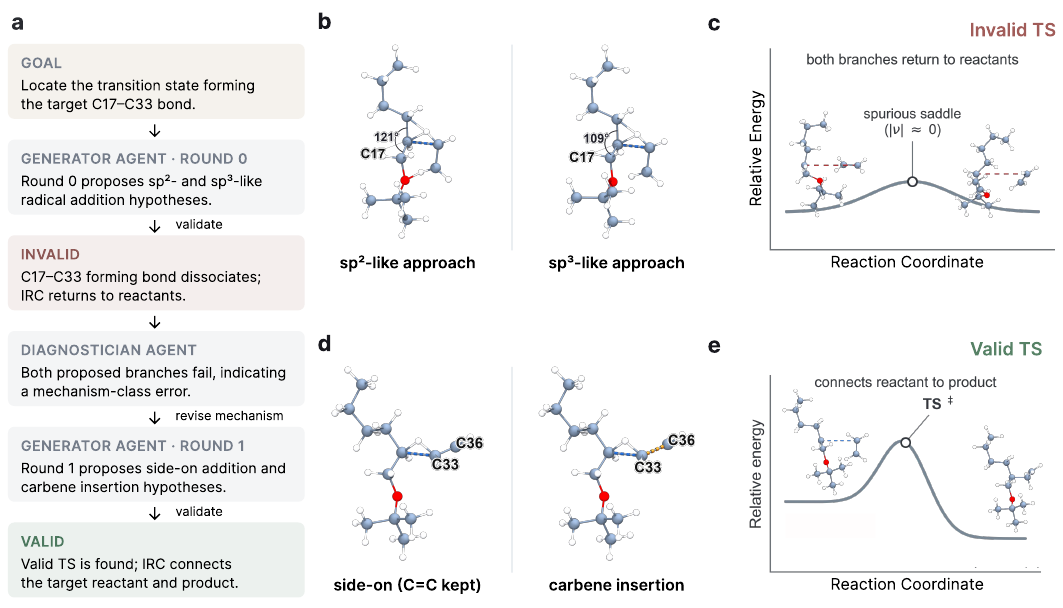}%
    \caption{\textbf{Mechanism replacement after diagnostic feedback on a
    carbene-addition reaction from the Complex elementary reactions subset.}
    \textbf{a}, Progression
    of the search: the round-0 radical-addition guesses fail validation, the
    diagnostician identifies a mechanism-class error and the round-1 mechanism
    replacement yields a validated TS. \textbf{b}, Round-0 TS guesses
    implementing the sp\textsuperscript{2}-like and sp\textsuperscript{3}-like
    radical-addition hypotheses. \textbf{c}, Schematic IRC profile of the
    round-0 candidates: both branches return to the separated reactants through
    a spurious saddle point with a near-zero imaginary frequency. \textbf{d},
    Round-1 TS guesses implementing the side-on (C=C retained) and
    carbene-insertion hypotheses. \textbf{e}, Schematic IRC profile of the
    validated round-1 TS, whose branches connect the target reactant and
    product.}
    \label{fig:case_study}
\end{figure*}

To make the endpoint-mismatch failure mode concrete, we examine one
representative carbene-addition reaction from the Complex elementary subset
(Fig.~\ref{fig:case_study}a). The mapped reaction forms a single new
C\textsubscript{17}--C\textsubscript{33} bond between a carbanion centre (C\textsubscript{17}) and a
carbene-like C\textsubscript{33}=C\textsubscript{36} fragment, so
the central mechanistic question is how the C\textsubscript{17} centre should approach the
C\textsubscript{33}=C\textsubscript{36} unit. In the initial generation round (round~0), the generator treated
this transformation as a concerted radical-addition-like process: both
proposed hypotheses placed C\textsubscript{17} in direct attack on
C\textsubscript{33} while keeping the C\textsubscript{33}=C\textsubscript{36}
bond largely intact, differing only in whether C\textsubscript{17} retained an
sp\textsuperscript{2}-like or adopted an sp\textsuperscript{3}-like local
geometry (Fig.~\ref{fig:case_study}b). All six round-0
guesses failed with a common structural signature: during TS optimisation the
forming C\textsubscript{17}--C\textsubscript{33} bond dissociated from
1.80--2.30\,\AA\ in the guesses to 3.85--4.93\,\AA, while the
C\textsubscript{33}=C\textsubscript{36} bond relaxed back to a double-bond-like
1.29--1.32\,\AA. Five of the six guesses converged to saddle points whose
imaginary frequencies were only $-14$ to $-96$\,cm$^{-1}$ and whose IRC
branches failed endpoint matching, in most cases returning the separated
reactants on both sides (Fig.~\ref{fig:case_study}c); the remaining guess did
not converge during TS optimisation yet showed the same dissociated forming bond.

The diagnostician interpreted this repeated collapse as a mechanism-class
failure rather than as a bond-length refinement problem, flagging both
round-0 mechanistic branches as requiring a new mechanism. It therefore abandoned
the one-centre radical-addition picture and proposed two alternative reaction
hypotheses: a side-on approach that brings
C\textsubscript{17} to the C\textsubscript{33}=C\textsubscript{36} unit while
preserving the double bond, and a carbene-insertion-like
pathway in which C\textsubscript{33} rehybridises towards
sp\textsuperscript{3} through partial breaking of the
C\textsubscript{33}=C\textsubscript{36} bond (Fig.~\ref{fig:case_study}d). In the first revision round (round~1), the generator converted these hypotheses into new TS guesses,
and one late-stage guess from each hypothesis passed strict validation.
The successful guesses had much larger reactive imaginary frequencies, near
$-1518$\,cm$^{-1}$, and their bidirectional IRC endpoints matched the
target reactant--product pair (Fig.~\ref{fig:case_study}e).

This case illustrates, at the single-reaction level, the recovery channel identified statistically above. Recovery came from replacing the mechanistic hypothesis that determined which
atoms participate in the reactive event, rather than from
fine-tuning a single forming-bond distance. The example therefore links the
endpoint-mismatch failure mode to the mechanism-replacement verdicts that drive
many successful revisions.

\section{Conclusions}\label{sec:conclusion}

Reaction-mechanism understanding is difficult to evaluate with text- or
connectivity-level criteria alone. A mechanistic hypothesis ultimately
claims that atoms can move through a specific three-dimensional pathway on a
potential energy surface, and that claim should be tested by physical criteria.
TSBench addresses this evaluation gap by using transition-state construction as
a mechanism-level probe of LLM agents. Given reactant and product information,
the agent must construct a three-dimensional TS guess; the guess is accepted only
when quantum-chemical validation optimises it to a saddle point whose
bidirectional IRC endpoints match the intended reactant--product pair.
This design turns reaction-mechanism understanding into a measurable task rather
than a qualitative judgement of whether a model gives a plausible explanation.

The resulting benchmark exposes a clear capability profile for current frontier
agents. Across 546 model--reaction evaluations, performance follows the chemical
regimes defined by the retained reactions: Simple elementary reactions are
broadly accessible, Intermediate elementary reactions separate model capabilities more
strongly, and the Complex elementary subset produces a shared low-ceiling regime.
The closed-loop diagnosis process is effective, raising aggregate strict success
from 50.4\% after the initial generation round to 66.8\% after two
revisions, and many successful revisions follow verdicts that replace the
proposed mechanism rather than only refine a local geometry. However, the same
validation pipeline also identifies the main remaining failure: endpoint
mismatch. Among unresolved evaluations with a determinable failure stage,
72.4\% terminate with IRC endpoints that do not match the target
reactant--product pair.

These results clarify both the promise and the limitation of LLM agents for
mechanism-sensitive chemistry. Current agents can often generate TS-like local
structures and can benefit from diagnostic feedback, but they still frequently
choose a reaction coordinate that does not connect the intended reactant and
product. This distinction matters for synthesis planning, catalyst screening,
process design and autonomous experimentation, where a plausible but incorrect
mechanism can lead to a wrong downstream decision. By coupling agentic structure
construction to quantum-chemical validation, TSBench provides a physics-grounded
measure of this missing capability. Because the validator is modular,
the same design extends naturally to higher-level electronic-structure methods,
solvated or catalytic reactions and multi-step mechanisms. Future progress
should therefore be measured by whether models reduce
endpoint-mismatch failures and more reliably formulate the mechanistic
hypotheses that make a TS guess physically valid, in addition to raising
aggregate success rates.

\section{Methods}\label{sec:methods}

\subsection{TSBench task definition}

Each TSBench instance specifies a target reaction by its reactant and product
SMILES, a three-dimensional reactant geometry, and the total charge and spin
multiplicity. The task input also includes reaction-centre annotations derived
from the reactant--product mapping, including formed and broken
bonds and bond-order changes. Given this symbolic
and geometric specification, the agent is required to submit one or more
three-dimensional TS guesses.

A model--reaction evaluation is counted as successful only if at least one
submitted guess passes the strict validation criterion and connects the target
reactant and product. At a high level, this criterion requires the candidate to
optimise to a first-order saddle point and for bidirectional IRC propagation to
recover endpoints whose connectivities match the target reactant--product pair.
The full validation pipeline and failure-stage assignment are described below.

\subsection{Benchmark construction}\label{sec:benchmark_construction}

TSBench was constructed from three previously published reaction datasets that
provide complementary elementary-reaction regimes. The first source is the
small-molecule test set of Zhang et al.~\cite{Zhang2025_RDA}, which was
originally used to evaluate reaction directional analysis--dimer (RDA-D) against
nudged elastic band (NEB) calculations. The second source is the textbook
elementary-reaction collection released with MDCD-NN~\cite{Li2025_MDCDNN}, which
covers familiar closed-shell and radical organic mechanism classes. The third
source is the BH9 barrier-height and reaction-energy benchmark~\cite{Prasad2022_BH9},
which contains more structurally and mechanistically diverse elementary
reactions, including rearrangements, cycloadditions, hydrogen-atom and proton
transfers, and nucleophilic additions.

We did not use the source datasets directly as benchmark instances. Instead,
candidate reactions from each source were first passed through the same ORCA--xTB
validation workflow used later to score agent-generated TS guesses. In this
screening step, each source-dataset reference TS was re-optimised at the GFN2-xTB
level, checked by numerical frequency analysis, propagated by bidirectional IRC
calculations, and accepted only when the optimised IRC endpoints reconstructed
the target reactant--product connectivity. Reactions whose reference structures
failed this strict validation test were excluded before constructing the final
benchmark. The reference structures were used only for this
compatibility screen and were withheld from the task inputs supplied to the
agent.

This xTB-based compatibility screen is necessary because TSBench evaluates agent
reasoning through an automated quantum-chemical pass/fail pipeline. If a
reference reaction cannot be reproduced by the chosen GFN2-xTB validator, then a
model could be penalised for a limitation of the electronic-structure workflow
rather than for an incorrect mechanistic hypothesis. Screening therefore makes
the benchmark internally consistent: every retained reaction has at least one
known TS that satisfies the same validation criteria applied to model-generated
structures.

\begin{figure*}[t]
    \centering
    \includegraphics[width=\textwidth,keepaspectratio]{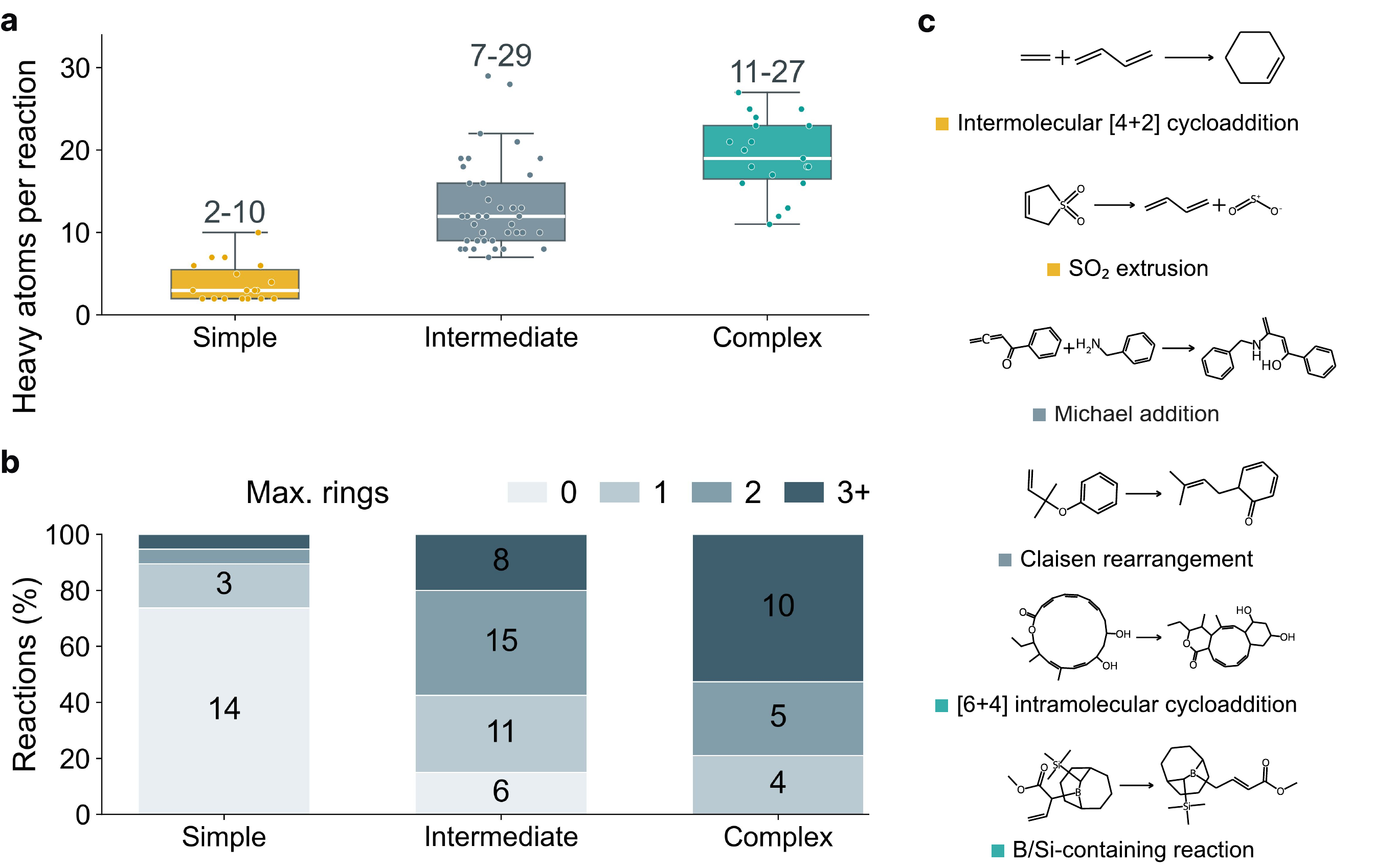}%
    \caption{\textbf{Composition of the 78-reaction TSBench corpus.}
    \textbf{a}, Heavy-atom counts of the retained reactions in
    the Simple elementary, Intermediate elementary and Complex elementary
    subsets after xTB-based compatibility screening. Boxes span the
    interquartile range, white centre lines mark the median, whiskers extend to
    the most extreme values within 1.5 times the interquartile range, points
    show individual reactions and the labels above each box give the
    minimum--maximum range. \textbf{b}, Distribution of
    the maximum number of rings present in each retained reaction, summarised by
    subset; numbers give reaction counts per ring class. \textbf{c}, Representative reactions from the three retained subsets;
    colours match the subset colours used in panel \textbf{a}.}
    \label{fig:dataset_overview}
\end{figure*}

After screening, the final TSBench corpus contains 78 reactions
(Fig.~\ref{fig:dataset_overview}a,b): 19 retained from the Zhang et al. test
systems, 40 retained from the textbook elementary-reaction collection, and 19
retained from BH9. The 40 retained textbook reactions comprise 20
neutral singlet and 20 neutral radical doublet reactions spanning 17 mechanism
labels, and the 19 retained BH9 reactions are stratified across ten BH9
reaction subclasses. Figure~\ref{fig:dataset_overview}a,b shows that these retained
reactions form a source- and structure-dependent complexity gradient. The Zhang
et al. reactions have the smallest molecular sizes and compact reaction centres,
so we refer to them as Simple elementary reactions. The retained textbook
reactions span broader organic mechanism families and moderate molecular sizes,
so we refer to them as Intermediate elementary reactions. The retained BH9
reactions include larger, charged, radical, ring-containing or multi-bond
reaction coordinates, so we refer to them as Complex elementary reactions.
Figure~\ref{fig:dataset_overview}c further shows representative reactions from
the three retained subsets, illustrating the progression from compact elementary
reactions to broader textbook mechanisms and more structurally demanding
BH9-derived reactions. Two-dimensional schemes of the 78 retained
reactions are provided in Supplementary Figs.~S1--S6.

\subsection{Agent tools and environment}

All evaluated models operated through the same fixed agent environment, executed
within Claude Code (v2.1.181). The agent roles, skills, prompt
templates, tool interfaces, workflow code and cross-round history retention were
held fixed across models. Verbatim prompt templates are provided in the Supplementary Information, and the complete skill definitions will be released with the benchmark code upon journal publication (see Code availability).

The atom-mapping tool was used during task preparation to align reactant and
product heavy atoms. Starting from the reactant and product SMILES, it generated
a heavy-atom mapping, re-indexed the mapping to the one-based atom numbering of
the reactant XYZ file, and derived the reaction-centre annotations supplied to
the agent: formed and broken bonds and bond-order changes.

The generator manipulated molecular structures through a structure-editing tool
designed to expose GaussView-like molecular editing operations to the agent~\cite{Dennington2016_GaussView}. Rather
than asking the model to rewrite XYZ coordinates directly, the tool let the agent
operate on a molecular geometry in the way a human would use an interactive
molecular editor: by selecting atoms or fragments and adjusting chemically
meaningful internal coordinates. The tool changed one bond length, angle or
dihedral per operation and could move either an atom or a connectivity-defined
fragment while holding the other atoms fixed. All operations used one-based atom
indices. A separate geometry-query tool measured bond lengths, bond angles and
dihedral angles, including comparisons with the starting reactant.
% After each editing operation, the generator protocol required the
% agent to re-measure the realised coordinate and to check retained bonds and
% short non-bonded contacts; intermediate builder structures were treated as
% working files rather than as independent TS guesses.
This design
tests whether the agent can reason about molecular shape and reaction geometry
while avoiding the low success rate associated with direct free-form XYZ editing.
AtomWorld similarly found that combining spatial reasoning with structure-file
syntax compounded task difficulty and argued that challenging structural
operations are better delegated to domain-specific tools in agentic
workflows~\cite{Lv2026_AtomWorld}.

\subsection{TS guess generation}

The generator converted the mapped reaction specification into candidate TS
structures (Fig.~\ref{fig:tsbench_workflow}b, left; Fig.~\ref{fig:tsbench_workflow}c).
For each round, it first interpreted the reaction centre and proposed alternative
mechanistic or conformational hypotheses. It then built three-dimensional
candidate geometries for those hypotheses. The generator operated under a fixed
TS-guess protocol that defined the available inputs, the output schema and the
permitted structure-editing workflow. It treated the mapped formed and broken
bonds as authoritative, inspected the starting geometry, and was requested to
propose two alternative mechanistic or conformational hypotheses per round. For
each hypothesis, it generated early-, middle- and late-stage guesses, giving a
nominal target of six structures per round.

\subsection{Quantum-chemical validation}

Every TS guess was evaluated independently by a validator that applied a fixed
quantum-chemical pipeline (Fig.~\ref{fig:tsbench_workflow}b, centre;
Fig.~\ref{fig:tsbench_workflow}c), using ORCA 6.1.1~\cite{Neese2025_ORCA6} interfaced to xTB 6.7.1~\cite{Bannwarth2021_xtb}. The pipeline comprised TS
optimisation, frequency analysis, bidirectional IRC propagation, endpoint
optimisation and connectivity-based endpoint matching. The ORCA \texttt{XTB} keyword
selects the GFN2-xTB Hamiltonian~\cite{Bannwarth2019_GFN2xTB}. Calculations were
performed without an implicit solvent model. The first stage used
\texttt{XTB OptTS TightOpt}, an initial calculated Hessian, Hessian
recalculation every five optimisation steps and a maximum of 200 optimisation
steps. A numerical frequency calculation was then performed at the optimised
geometry with the xTB accuracy parameter set to 1.0.

Candidates with exactly one imaginary frequency were propagated in both
directions with an IRC calculation using \texttt{HessMode 0} and a maximum of
100 iterations. The terminal structure from each IRC branch was subsequently
geometry-optimised at the same GFN2-xTB level. RDKit~\cite{Landrum2024_RDKit} was used to convert each
optimised endpoint into a canonical SMILES with stereochemistry
removed. The two observed endpoint SMILES and
the target reactant--product SMILES were each sorted before comparison, so either
IRC direction could lead to either endpoint.

A guess was counted as a strict success only if TS optimisation terminated and
converged, the frequency calculation terminated with exactly one imaginary
frequency, the bidirectional IRC and both endpoint optimisations completed, and
the unordered pair of endpoint SMILES exactly matched the target reactant and
product. The first unmet condition assigned the failure stage as guess
preparation, TS optimisation, frequency analysis, IRC propagation, endpoint
optimisation or endpoint matching.

\subsection{Failure diagnosis and iterative revision}

\begin{figure}[p]
    \centering
    \includegraphics[width=\textwidth,height=0.6\textheight,keepaspectratio]{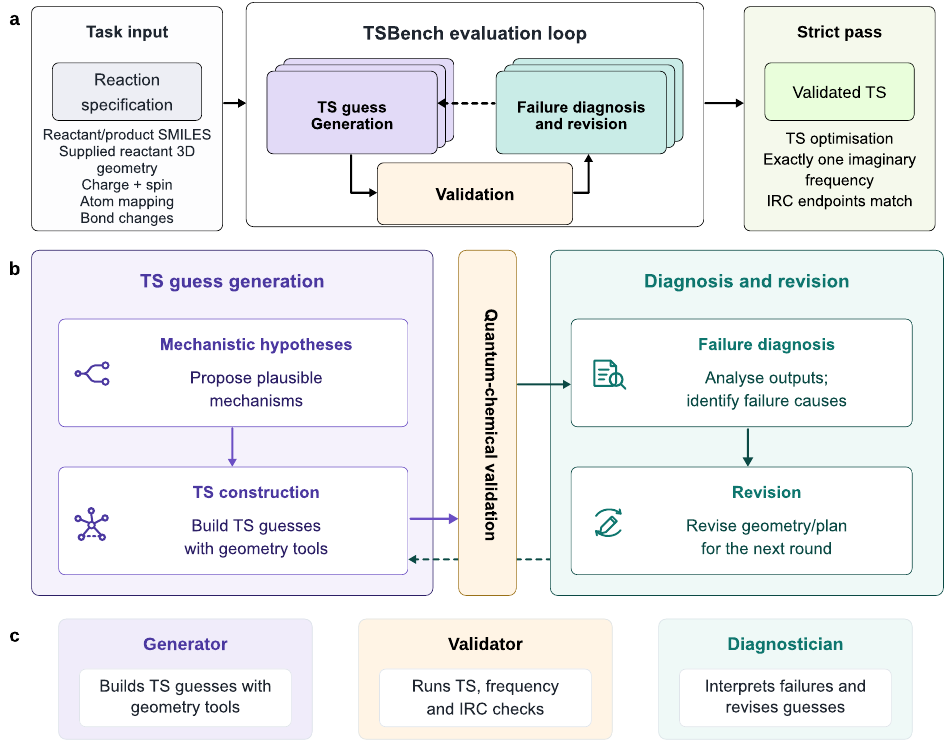}
    \caption{\textbf{Physics-grounded TSBench evaluation workflow.}
    \textbf{a}, Overview of the closed generate--validate--diagnose--revise loop.
    A reaction specification---reactant and product SMILES, a three-dimensional
    reactant geometry, charge and multiplicity, and the derived heavy-atom map and
    bond changes---enters an evaluation loop in which TS-guess generation,
    quantum-chemical validation and failure diagnosis alternate until a candidate
    strictly passes (converged TS optimisation, a single imaginary frequency and
    IRC endpoints matching the target) and is returned as a validated TS.
    \textbf{b}, Detailed view of one round, matching the three workflow modules expanded in Methods:
    TS guess generation (left), quantum-chemical validation (centre) and failure
    diagnosis with iterative revision (right). The generator interprets the
    reaction, proposes two mechanistic hypotheses and uses structure-editing tools
    to build TS guesses; each guess is checked by the quantum-chemical validation
    pipeline; when every guess in a round fails, the diagnostician analyses the
    calculation outputs and returns a mechanism- or geometry-level revision that
    seeds the next round.
    \textbf{c}, Agent environment used for all model evaluations. The fixed roles
    are a generator that builds three-dimensional TS guesses from the mapped bond
    changes using geometry-query and structure-editing tools, a validator that
    runs TS optimisation, frequency analysis, bidirectional IRC, endpoint
    optimisation and SMILES matching, and a diagnostician that interprets failed
    calculations and IRCs and recommends geometry refinement or mechanism
    replacement.}
    \label{fig:tsbench_workflow}
\end{figure}

When no TS guess structures in a round passed strict validation, the agent
invoked a fixed diagnostician protocol (Fig.~\ref{fig:tsbench_workflow}b, right;
Fig.~\ref{fig:tsbench_workflow}c) to analyse the failed branches. It parsed the
ORCA outputs, compared key internal coordinates in the initial and optimised
structures, inspected the IRC trajectory when available and compared the
reconstructed endpoint identities. The diagnostician first checked whether an
initial guess contained corrupted connectivity or short contacts before
attributing failure to the proposed mechanism. It then returned a structured
diagnosis, an acceptance verdict, a branch decision, numerical suggestions
for subsequent builder operations and, when it identified a mechanism-class failure, recommended alternative mechanistic directions for the next round.

The diagnosis was returned to the generator, closing the round-to-round loop
(Fig.~\ref{fig:tsbench_workflow}a,b). The generator could refine the geometry of
a continuing branch or replace a mechanistic hypothesis when multiple chemically
reasonable guesses showed a common failure pattern. An evaluation stopped as soon as
any guess achieved strict success.

\subsection{Model evaluation}

We evaluated GLM-5.1, GPT-5.4, MiniMax-M2.7, Claude Sonnet 4.6, DeepSeek V4 Pro,
Gemini 3.1 Pro Preview and Qwen 3.6 Plus. Each model--reaction pair defined one
evaluation, giving 546 evaluations in total. Each model--reaction
combination was evaluated once. Each evaluation was allowed a
maximum of three generate--validate--diagnose--revise rounds.

For a model--reaction evaluation $i$, strict success at round $k$ was defined as
an indicator that at least one guess had passed the complete validator by the
end of round $k$. Cumulative success at round $k$ was the mean of this indicator
over the relevant set of evaluations. We report this quantity by model, reaction
subset and revision round. For unsuccessful evaluations, the terminal failure
stage was defined as the most frequent failure stage among failed guesses in
the final non-empty round. Ties were resolved by the order of the validation
pipeline: TS optimisation, frequency analysis, IRC propagation, endpoint
optimisation and endpoint matching.

\backmatter

\bmhead{Supplementary information}

The Supplementary Information contains subset-resolved cumulative
success tables (Supplementary Tables~S1--S3), failure-stage breakdowns
(Supplementary Tables~S4 and S5), two-dimensional schemes of the 78
benchmark reactions (Supplementary Figs.~S1--S6), diagnostic-feedback statistics
(Supplementary Figs.~S7 and S8) and Supplementary Methods, including verbatim
agent prompts and the workflow-record schema.

% Acknowledgements (funding sources) to be added before journal submission.

\bmhead{Data availability}

The TSBench reaction corpus, per-evaluation search records and
aggregated result files will be made publicly available upon
journal publication, and are available from the authors upon reasonable request
in the meantime. The source reaction datasets are available from the
original publications~\cite{Zhang2025_RDA,Li2025_MDCDNN,Prasad2022_BH9}.

\bmhead{Code availability}

The benchmark runner, agent skill definitions, geometry tools and the
quantum-chemical validation workflow will be made publicly available
upon journal publication, and are available from the authors upon reasonable
request in the meantime.

\bmhead{Author contributions}

X.X. designed the benchmark, implemented the agent environment and
validation pipeline, performed the evaluations and analysed the data. T.Z.
supervised the project. Both authors wrote and revised the manuscript. 

\bmhead{Competing interests}

The authors declare no competing interests.

\bibliography{sn-bibliography}

\clearpage

\end{document}

\maketitle

\section{Benchmark reaction schemes}

Supplementary Figs.~S1--S6 show two-dimensional reactant--product schemes for
the 78 TSBench reactions: the Simple elementary subset in Fig.~S1, the
Intermediate elementary subset in Figs.~S2--S4 and the Complex elementary
subset in Figs.~S5 and S6. Schemes are numbered within each subset, following
the order of the reaction records used in the benchmark;
repeated components are collapsed with stoichiometric coefficients.
Structures are drawn from the benchmark input SMILES, which the evaluated
agents received verbatim. Automated bond-order perception encodes open-shell
species as closed-shell, charge-separated strings; such species are depicted
here as the corresponding physical radical or carbene structures whenever
that reconstruction reproduces the recorded system charge and spin
multiplicity exactly, and the verbatim encoding is drawn otherwise (for
example for carbon monoxide and diazomethane, whose charge-separated SMILES
are standard closed-shell representations).

\begin{figure}[p]
\centering
\includegraphics[width=\textwidth,height=0.88\textheight,keepaspectratio]{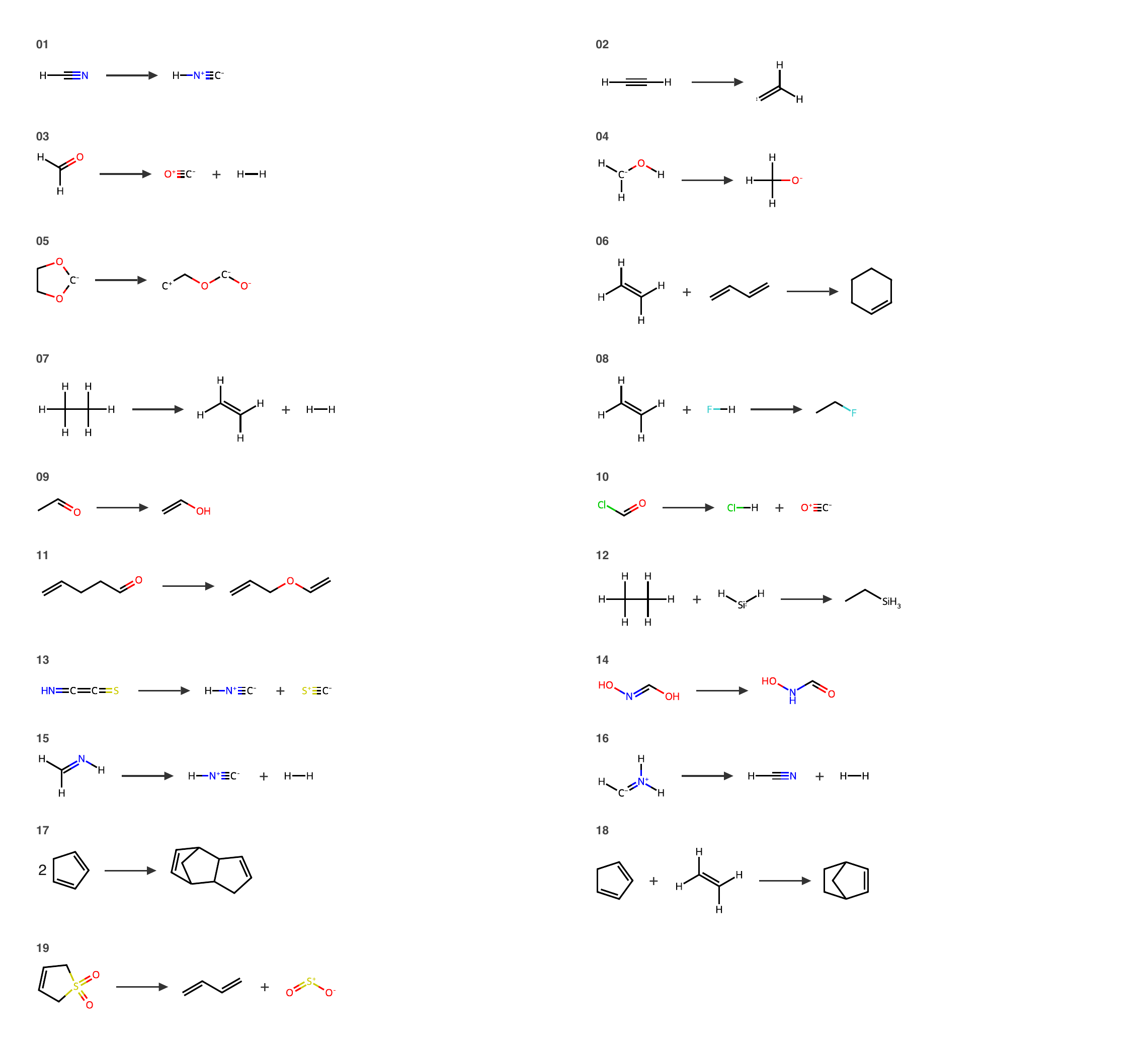}
\caption{Two-dimensional reactant--product schemes of the 19 Simple
elementary reactions.}
\label{fig:atlas_simple}
\end{figure}

\begin{figure}[p]
\centering
\includegraphics[width=\textwidth,height=0.88\textheight,keepaspectratio]{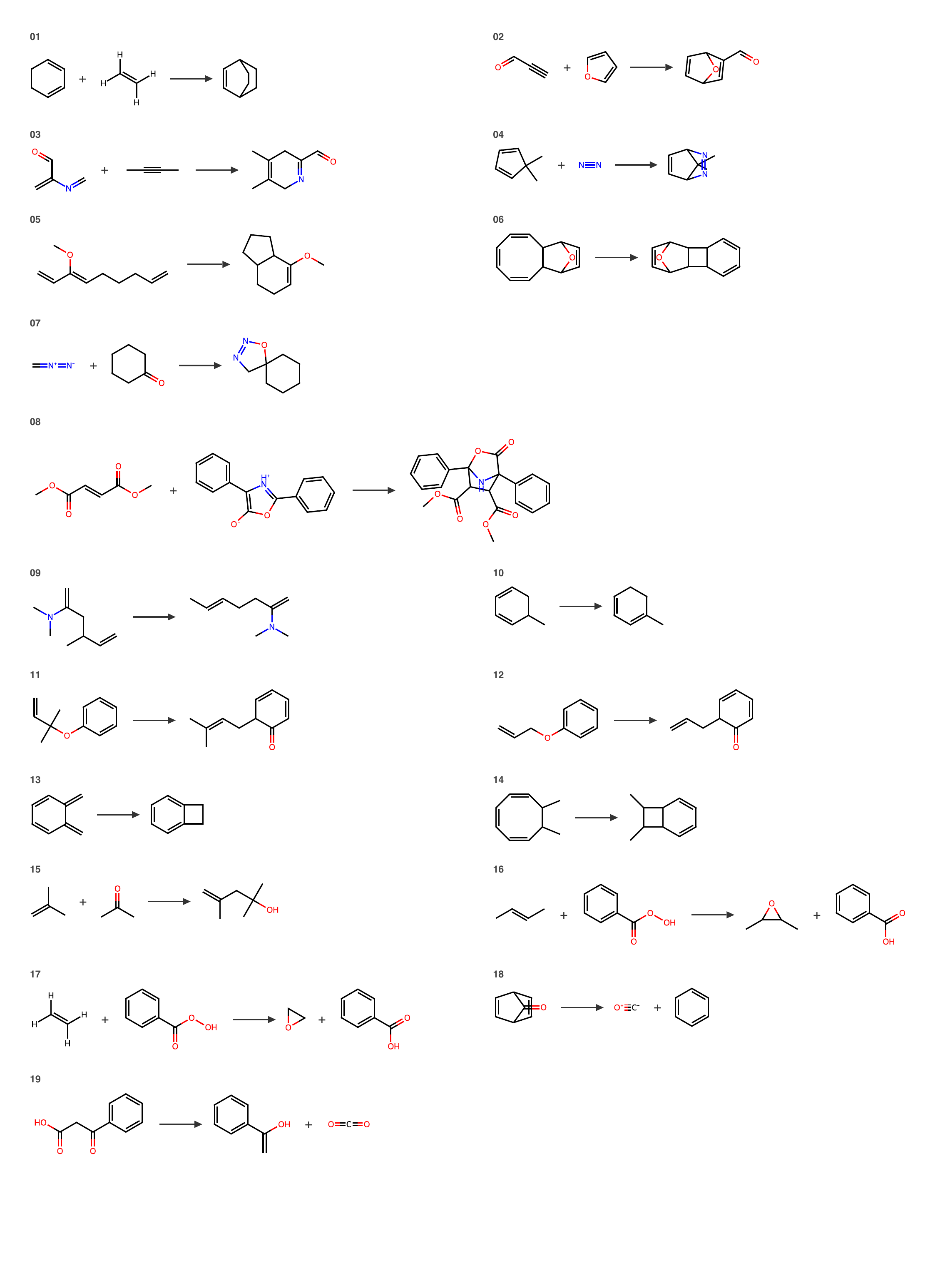}
\caption{Two-dimensional reactant--product schemes of the Intermediate
elementary reactions (part 1 of 3, reactions 1--19).}
\label{fig:atlas_intermediate_1}
\end{figure}

\begin{figure}[p]
\centering
\includegraphics[width=\textwidth,height=0.88\textheight,keepaspectratio]{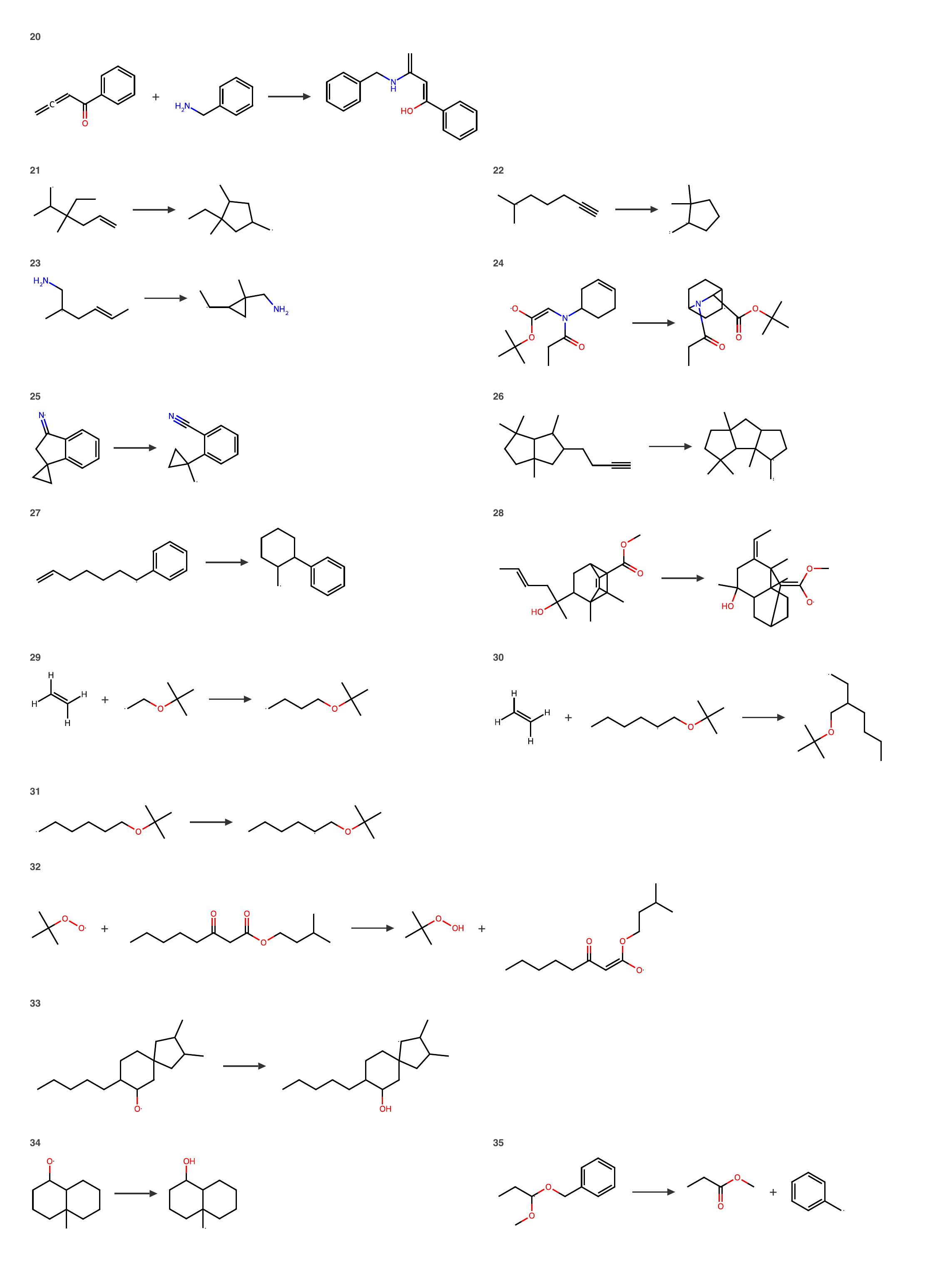}
\caption{Two-dimensional reactant--product schemes of the Intermediate
elementary reactions (part 2 of 3, reactions 20--35).}
\label{fig:atlas_intermediate_2}
\end{figure}

\begin{figure}[p]
\centering
\includegraphics[width=\textwidth,height=0.88\textheight,keepaspectratio]{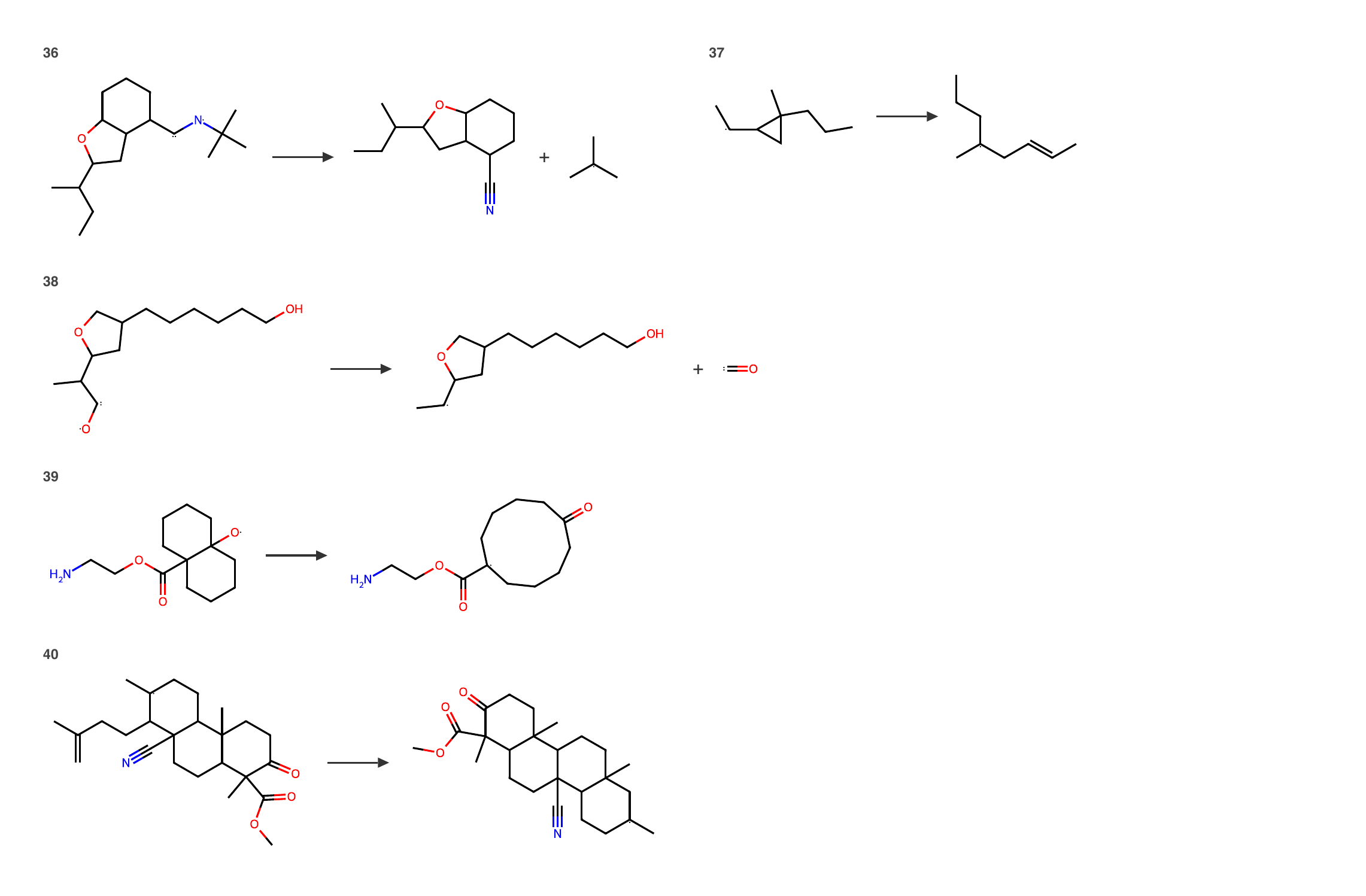}
\caption{Two-dimensional reactant--product schemes of the Intermediate
elementary reactions (part 3 of 3, reactions 36--40).}
\label{fig:atlas_intermediate_3}
\end{figure}

\begin{figure}[p]
\centering
\includegraphics[width=\textwidth,height=0.88\textheight,keepaspectratio]{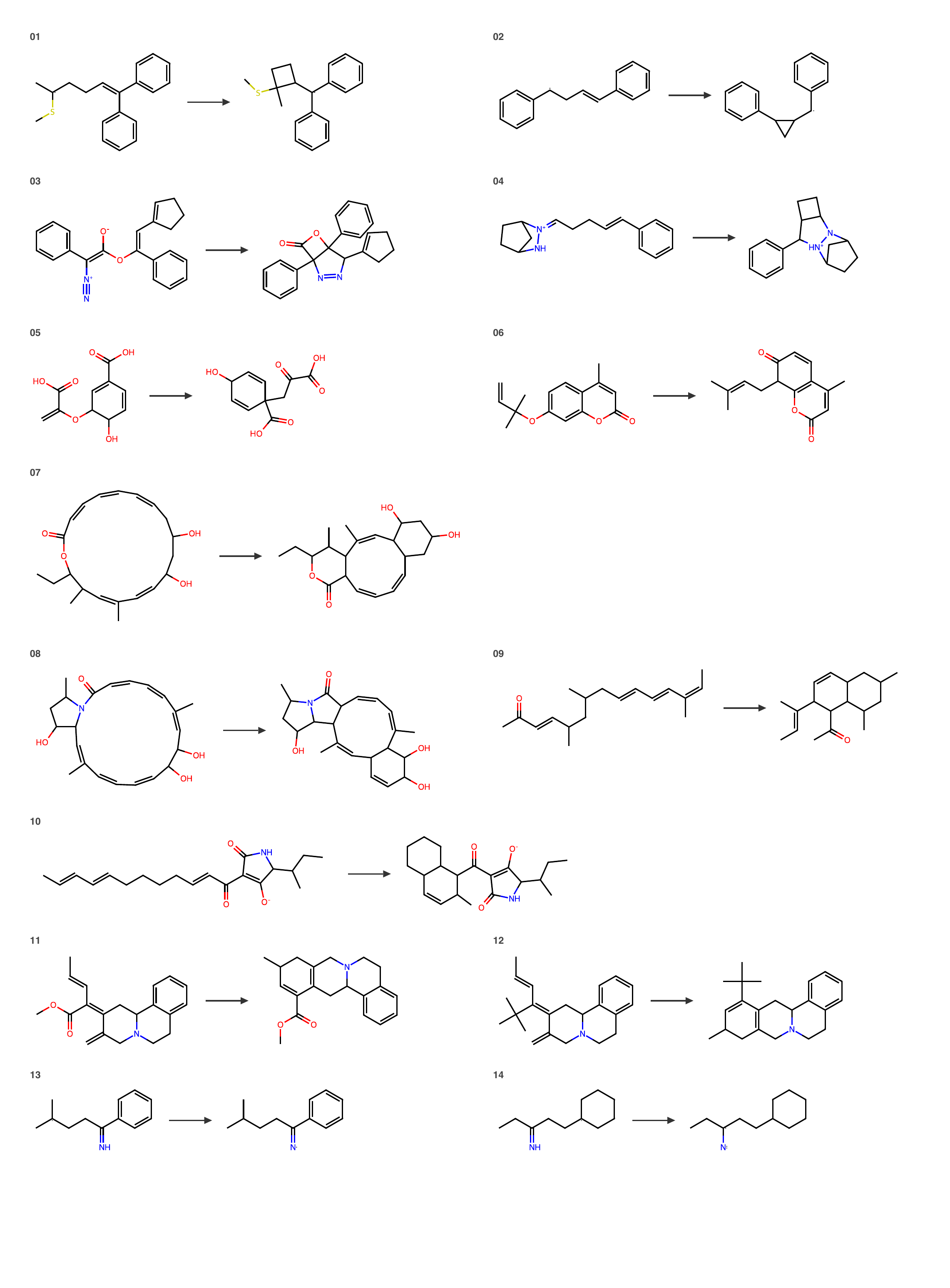}
\caption{Two-dimensional reactant--product schemes of the Complex elementary
reactions (part 1 of 2, reactions 1--14).}
\label{fig:atlas_complex_1}
\end{figure}

\begin{figure}[p]
\centering
\includegraphics[width=\textwidth,height=0.88\textheight,keepaspectratio]{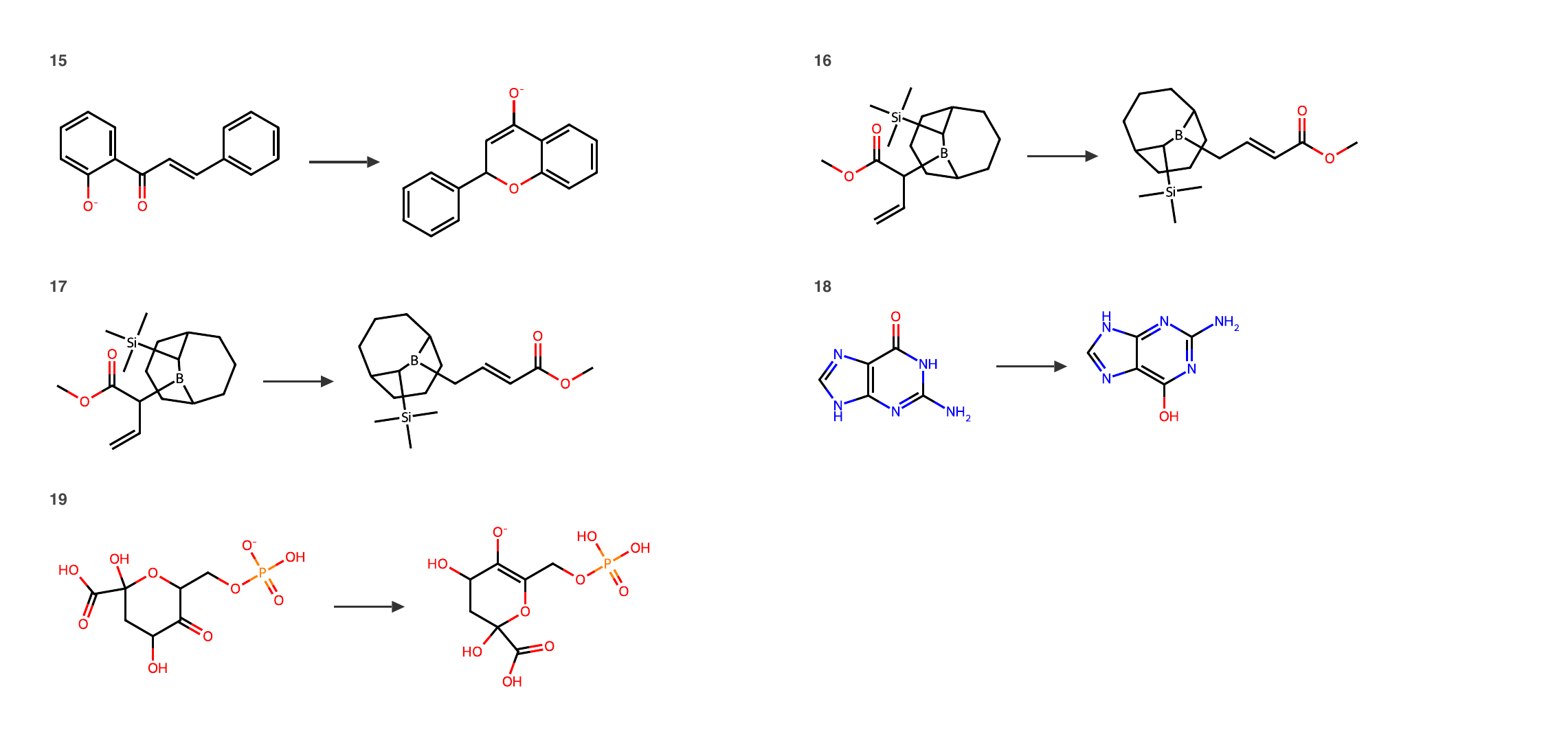}
\caption{Two-dimensional reactant--product schemes of the Complex elementary
reactions (part 2 of 2, reactions 15--19).}
\label{fig:atlas_complex_2}
\end{figure}

\clearpage

\section{Supplementary Results}

Subset names follow the main text: Simple elementary reactions denote compact
small-molecule systems from the Zhang et al. test systems~\cite{Zhang2025_RDA},
Intermediate elementary reactions denote textbook organic mechanisms of
moderate molecular size from the collection released with
MDCD-NN~\cite{Li2025_MDCDNN}, and Complex elementary reactions denote larger or
more topologically demanding reaction coordinates from the BH9
benchmark~\cite{Prasad2022_BH9}. All values use the three-round
evaluation budget defined in the main text (one initial generation round
followed by up to two diagnosis-driven revision rounds).

\begin{table}[ht]
\centering
\caption{Cumulative TS-search success rate (\%) across revision rounds for the
Simple elementary reactions subset. Values in parentheses are the gain over the
previous round.}
\label{tab:cumulative_gas_phase}
\small
\setlength{\tabcolsep}{6pt}
\begin{tabular}{lccc}
\toprule
Model & Initial & Revision 1 & Revision 2 \\
\midrule
GPT-5.4 & 63.2 & 78.9 (+15.8) & 78.9 (+0.0) \\
Claude Sonnet 4.6 & 78.9 & 89.5 (+10.5) & 89.5 (+0.0) \\
Gemini 3.1 Pro Preview & 68.4 & 84.2 (+15.8) & 84.2 (+0.0) \\
GLM-5.1 & 73.7 & 78.9 (+5.3) & 78.9 (+0.0) \\
DeepSeek V4 Pro & 63.2 & 73.7 (+10.5) & 84.2 (+10.5) \\
MiniMax-M2.7 & 36.8 & 63.2 (+26.3) & 73.7 (+10.5) \\
Qwen 3.6 Plus & 42.1 & 63.2 (+21.1) & 73.7 (+10.5) \\
\bottomrule
\end{tabular}
\end{table}

\begin{table}[ht]
\centering
\caption{Cumulative TS-search success rate (\%) across revision rounds for the
Intermediate elementary reactions subset. Values in parentheses are the gain
over the previous round.}
\label{tab:cumulative_textbook_organic}
\small
\setlength{\tabcolsep}{6pt}
\begin{tabular}{lccc}
\toprule
Model & Initial & Revision 1 & Revision 2 \\
\midrule
GPT-5.4 & 52.5 & 52.5 (+0.0) & 55.0 (+2.5) \\
Claude Sonnet 4.6 & 67.5 & 75.0 (+7.5) & 90.0 (+15.0) \\
Gemini 3.1 Pro Preview & 57.5 & 67.5 (+10.0) & 72.5 (+5.0) \\
GLM-5.1 & 72.5 & 77.5 (+5.0) & 85.0 (+7.5) \\
DeepSeek V4 Pro & 55.0 & 70.0 (+15.0) & 72.5 (+2.5) \\
MiniMax-M2.7 & 45.0 & 62.5 (+17.5) & 62.5 (+0.0) \\
Qwen 3.6 Plus & 50.0 & 65.0 (+15.0) & 65.0 (+0.0) \\
\bottomrule
\end{tabular}
\end{table}

\begin{table}[ht]
\centering
\caption{Cumulative TS-search success rate (\%) across revision rounds for the
Complex elementary reactions subset. Values in parentheses are the gain over the
previous round.}
\label{tab:cumulative_mechanistically_complex}
\small
\setlength{\tabcolsep}{6pt}
\begin{tabular}{lccc}
\toprule
Model & Initial & Revision 1 & Revision 2 \\
\midrule
GPT-5.4 & 26.3 & 36.8 (+10.5) & 36.8 (+0.0) \\
Claude Sonnet 4.6 & 21.1 & 31.6 (+10.5) & 36.8 (+5.3) \\
Gemini 3.1 Pro Preview & 15.8 & 31.6 (+15.8) & 42.1 (+10.5) \\
GLM-5.1 & 31.6 & 42.1 (+10.5) & 47.4 (+5.3) \\
DeepSeek V4 Pro & 31.6 & 42.1 (+10.5) & 52.6 (+10.5) \\
MiniMax-M2.7 & 26.3 & 31.6 (+5.3) & 36.8 (+5.3) \\
Qwen 3.6 Plus & 26.3 & 36.8 (+10.5) & 47.4 (+10.5) \\
\bottomrule
\end{tabular}
\end{table}

% Generated by scripts/plot_record.py; do not edit manually.
\begin{table}[ht]
\centering
\caption{Failure-stage distribution for failed TS guesses by reaction subset. Subset labels use the short names defined at the start of this section. Counts follow the plotting definition used for the main-text failure-stage figure: input-preparation failures are excluded and failed guesses are aggregated over the three-round evaluation budget defined in the main text. Percentages are relative to failed TS guesses within each subset.}
\label{tab:failure_step_breakdown}
\small
\setlength{\tabcolsep}{4pt}
\resizebox{\textwidth}{!}{%
\begin{tabular}{L{4.0cm}r r r r}
\toprule
Subset & Failed guesses & Endpoint mismatch, n (\%) & TS optimisation, n (\%) & Other validation stages, n (\%) \\
\midrule
Simple & 879 & 579 (65.9) & 212 (24.1) & 88 (10.0) \\
Intermediate & 2217 & 1374 (62.0) & 531 (24.0) & 312 (14.1) \\
Complex & 1516 & 1137 (75.0) & 267 (17.6) & 112 (7.4) \\
\bottomrule
\end{tabular}
}%
\end{table}

% Generated by scripts/plot_record.py; do not edit manually.
\begin{table}[ht]
\centering
\caption{Model-resolved failure-stage distribution for failed TS guesses. Subset group labels use the short names defined at the start of this section. Counts use the same plotting definition as Supplementary Table~\ref{tab:failure_step_breakdown}: input-preparation failures are excluded and failed guesses are aggregated over the three-round evaluation budget defined in the main text. Percentages are relative to failed TS guesses for the corresponding subset--model pair.}
\label{tab:failure_step_breakdown_by_model}
\small
\setlength{\tabcolsep}{4pt}
\resizebox{\textwidth}{!}{%
\begin{tabular}{L{3.2cm}r r r r}
\toprule
Model & Failed guesses & Endpoint mismatch, n (\%) & TS optimisation, n (\%) & Other validation stages, n (\%) \\
\midrule
\multicolumn{5}{l}{\textit{Simple}} \\
GPT-5.4 & 134 & 88 (65.7) & 32 (23.9) & 14 (10.4) \\
Claude Sonnet 4.6 & 100 & 60 (60.0) & 28 (28.0) & 12 (12.0) \\
Gemini 3.1 Pro Preview & 86 & 56 (65.1) & 25 (29.1) & 5 (5.8) \\
GLM-5.1 & 113 & 85 (75.2) & 20 (17.7) & 8 (7.1) \\
DeepSeek V4 Pro & 140 & 92 (65.7) & 39 (27.9) & 9 (6.4) \\
MiniMax-M2.7 & 184 & 127 (69.0) & 41 (22.3) & 16 (8.7) \\
Qwen 3.6 Plus & 122 & 71 (58.2) & 27 (22.1) & 24 (19.7) \\
\addlinespace
\multicolumn{5}{l}{\textit{Intermediate}} \\
GPT-5.4 & 393 & 246 (62.6) & 98 (24.9) & 49 (12.5) \\
Claude Sonnet 4.6 & 265 & 167 (63.0) & 46 (17.4) & 52 (19.6) \\
Gemini 3.1 Pro Preview & 268 & 151 (56.3) & 81 (30.2) & 36 (13.4) \\
GLM-5.1 & 259 & 158 (61.0) & 57 (22.0) & 44 (17.0) \\
DeepSeek V4 Pro & 348 & 229 (65.8) & 80 (23.0) & 39 (11.2) \\
MiniMax-M2.7 & 386 & 237 (61.4) & 98 (25.4) & 51 (13.2) \\
Qwen 3.6 Plus & 298 & 186 (62.4) & 71 (23.8) & 41 (13.8) \\
\addlinespace
\multicolumn{5}{l}{\textit{Complex}} \\
GPT-5.4 & 236 & 193 (81.8) & 30 (12.7) & 13 (5.5) \\
Claude Sonnet 4.6 & 249 & 205 (82.3) & 33 (13.3) & 11 (4.4) \\
Gemini 3.1 Pro Preview & 163 & 122 (74.8) & 36 (22.1) & 5 (3.1) \\
GLM-5.1 & 228 & 168 (73.7) & 38 (16.7) & 22 (9.6) \\
DeepSeek V4 Pro & 230 & 169 (73.5) & 37 (16.1) & 24 (10.4) \\
MiniMax-M2.7 & 232 & 154 (66.4) & 52 (22.4) & 26 (11.2) \\
Qwen 3.6 Plus & 178 & 126 (70.8) & 41 (23.0) & 11 (6.2) \\
\bottomrule
\end{tabular}
}%
\end{table}

\clearpage
\subsection{Diagnostic-feedback statistics}

\begin{figure}[ht]
\centering
\includegraphics[width=0.72\textwidth,keepaspectratio]{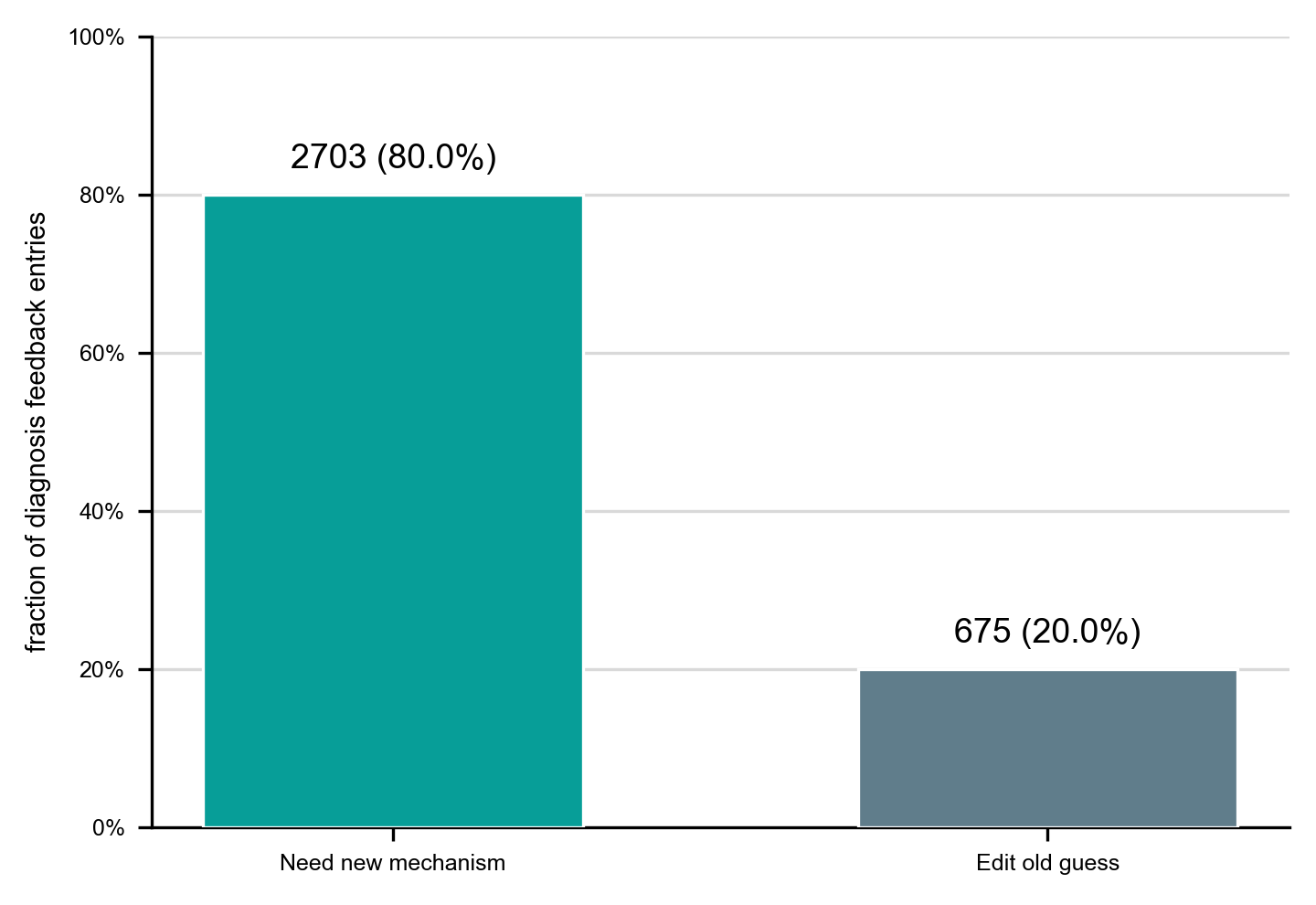}
\caption{Fraction of classifiable diagnostic feedback entries that request a new
mechanism versus editing the old guess. Counts aggregate
\texttt{if\_need\_new\_mechanism} annotations across the three subsets within
the three-round evaluation budget defined in the main text.}
\label{fig:diagnosis_need_new_mechanism}
\end{figure}

\begin{figure}[ht]
\centering
\includegraphics[width=0.72\textwidth,keepaspectratio]{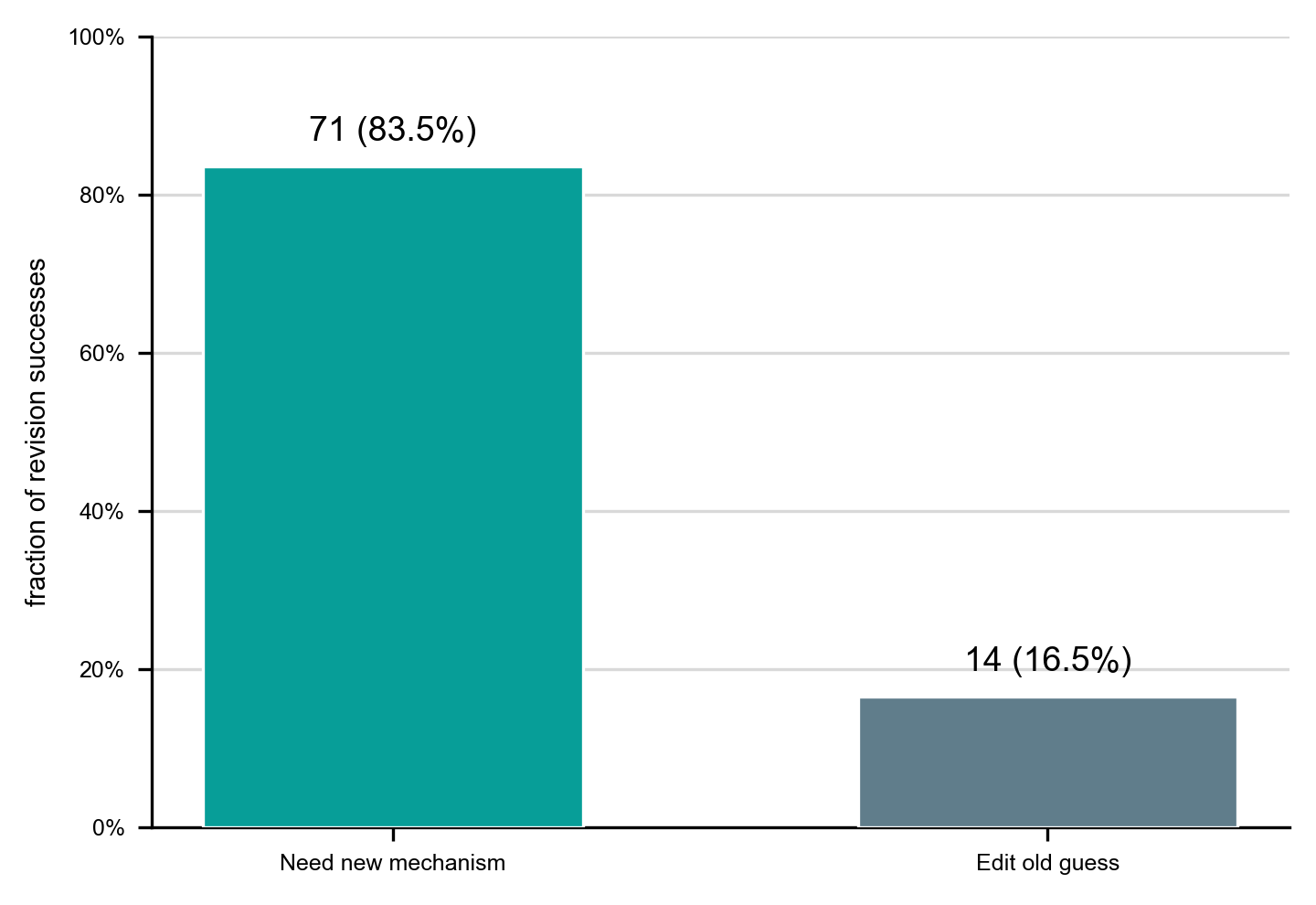}
\caption{Successful revisions grouped by the preceding diagnosis. Only
model--reaction evaluations that first succeeded in revision round 1 or 2 and
whose preceding diagnosis contained a classifiable
\texttt{if\_need\_new\_mechanism} annotation are included: of the 90
revision-round successes, 85 had a classifiable preceding annotation.
Initial-round successes and successes without a classifiable preceding
annotation are excluded.}
\label{fig:success_after_previous_diagnosis}
\end{figure}

\clearpage

\section{Supplementary Methods}

The benchmark runner invoked one generator call per round and, when no
submitted guess passed validation, one diagnostician call for that round. The
default single-round prompt templates passed to the Claude Code harness are
shown below.
Placeholders enclosed in braces were filled from the reaction inputs, reaction
mapping and current workflow state at runtime.

\noindent\textbullet\quad\textbf{Prompt for TS-guess generation}

\begin{lstlisting}[style=promptbox]
Generate TS guesses for exactly one round.

Use the repository skill `/generate-ts-guesses`.

Round: {round}
Reaction ID: {reaction_id}
Record JSON: {record_path}
Workdir: {workdir}
Reactant SMILES: {reactant_smiles}
Product SMILES: {product_smiles}
Reactant XYZ: {reactant_xyz}

Full reaction_mapping.json:
```json
{reaction_mapping_json}
```

Requirements:
- Do not run the full TS search coordinator.
- Only do this single generation task.
- Read `{record_path}` as the full record JSON array.
- Use the last array element as the current round block.
- Previous array elements contain prior guesses and diagnosis_feedback.
- If previous diagnosis_feedback has `if_need_new_mechanism: true`, abandon those failed mechanisms and generate fundamentally different mechanism hypotheses instead of only editing the old geometry.
- Treat `branch_decision: stop_this_branch` as stopping that specific failed branch, not as stopping the whole TS search when new-mechanism feedback exists.
- Do not manually edit or rewrite the full `{record_path}` file.
- Write the generated guesses list to `{round_dir}/guesses.record_update.json`.
- Then run `python tools/record_update.py set-guesses --record {record_path} --json-file {round_dir}/guesses.record_update.json`.
- Do not create generator_output.round_N.json.
- Keep all generated guess files under `{round_dir}`.
- Return a short status message with the record path.
\end{lstlisting}

\noindent\textbullet\quad\textbf{Prompt for TS-failure diagnosis}

\begin{lstlisting}[style=promptbox]
Diagnose TS validation failures for exactly one round.

Use the repository skill `/diagnose-ts-failures`.

Round: {round}
Reaction ID: {reaction_id}
Record JSON: {record_path}
Workdir: {workdir}

Full reaction_mapping.json:
```json
{reaction_mapping_json}
```

Requirements:
- Do not run the full TS search coordinator.
- Only do this single diagnosis task.
- Read `{record_path}` as the full record JSON array.
- Use the last array element as the current round block.
- Diagnose the current block's guesses using their validation fields.
- Do not manually edit or rewrite the full `{record_path}` file.
- Write the diagnosis feedback list to `{round_dir}/diagnosis_feedback.record_update.json`.
- Then run `python tools/record_update.py set-diagnosis-feedback --record {record_path} --json-file {round_dir}/diagnosis_feedback.record_update.json`.
- Do not create diagnostician_output.round_N.json.
- Return a short status message with the record path.
\end{lstlisting}

The complete generator and diagnostician skill texts, together with the
runner, validator, atom-mapping and geometry-tool sources, will be made
publicly available upon journal publication and are available from the
authors upon reasonable request in the meantime (see Code availability
in the main text). These skill files define the
fixed TS-guess schema, the geometry-tool protocol, the diagnostic-feedback
schema and the rule that new-mechanism feedback should trigger replacement of a
failed mechanistic hypothesis rather than only coordinate editing.

Each \texttt{record.json} file is an array of round objects, and each round
contains only \texttt{generation\_policy}, \texttt{guesses} and
\texttt{diagnosis\_feedback}. Generated guess entries were first written with pending validation
status; after quantum-chemical validation, the same entries were populated with
the validation status, failed step, file paths and endpoint-matching metrics.
Diagnostic feedback was then written into the same round block and included the
failure interpretation, branch decision, \texttt{if\_need\_new\_mechanism}
annotation and suggested builder operations. The record-update tool validated
generated JSON fragments and wrote the current round atomically. The complete
record array was passed to subsequent generator and diagnostician calls so that
revision could use all earlier structures, validation outcomes and diagnoses.

\bibliographystyle{sn-nature}
\bibliography{sn-bibliography}